\documentclass[journal]{IEEEtran}
\usepackage{tabularx}
\usepackage{subfigure} 
\usepackage{dblfloatfix,fixltx2e}
\usepackage{mathtools}
\usepackage{tabularx}
\usepackage{xcolor}

\newcommand{\Fig}[1]{Figure~\ref{fig:#1}}
\newcommand{\Sec}[1]{Section~\ref{sec:#1}}

\newcommand{\Eq}[1]{(\ref{eq:#1})}

\newcommand{\Ac}{\mathcal{A}}
\newcommand{\Zc}{\mathcal{Z}}
\newcommand{\Dc}{\mathcal{D}}
\newcommand{\Lc}{\mathcal{L}}
\newcommand{\Rc}{\mathcal{R}}
\newcommand{\Kc}{\mathcal{K}}

\begin{document}

\title{Planning UAV Activities for Efficient User Coverage in Disaster Areas}
\author{F. Malandrino, C.-F. Chiasserini, C. Casetti, L. Chiaraviglio, A. Senacheribbe\thanks{This work is licensed under Creative Commons License CC-BY-NC-ND.}}
\maketitle
\begin{abstract}

Climate changes brought about by global warming as well as man-made environmental
changes are often the cause of sever natural disasters. ICT, which is itself responsible
for global warming due to its high carbon footprint, can play a role in alleviating the 
consequences of such hazards by providing reliable, resilient means of communication 
during  a disaster crisis. In this paper, we explore the provision of wireless
coverage through UAVs (Unmanned Aerial Vehicles) to complement, or replace, the traditional
communication infrastructure. The use of UAVs is indeed crucial in emergency scenarios, as they allow for the quick and easy  deployment of micro and pico cellular base stations where needed.  We characterize the movements of UAVs and define 
an optimization problem to determine the best UAV coverage that maximizes the user throughput,
while maintaining fairness across the different parts of the geographical area that has been affected by the disaster. To evaluate our strategy,
we simulate a flooding in San Francisco and the car traffic resulting from people 
seeking safety on higher ground.
\end{abstract}

\section{Introduction}

Every second, one person in the world has her or his life irremediably 
affected by natural disasters \cite{report_disaster}. As climate and environmental changes become more and
more pronounced, they will likely lead to more frequent, severe, natural hazards, with heavy impact on 
peoples' welfare and nations' economies \cite{van2006impacts}.
Therefore, it is important to use the latest communication technologies to make public safety services as effective as possible \cite{smith2009technology}.

Relevant scenarios include the natural disasters such as earthquakes,
fires or floods that disrupt traditional communication infrastructures,
or that displace large crowds, overloading what is left of the
communication infrastructure \cite{gomez2014enabling}. In these scenarios, people are in need of
directions, advice, updates on the current situation and they often need
to send out requests for help or to make their location known to rescue
teams. Ideally, users should be able to communicate directly between
themselves \cite{shklovski2008finding,ali2016architecture}, even in the absence of a cellular infrastructure - a common
occurrence in disaster scenarios. Thus, it is crucial that
solutions and platforms are developed to integrate, or thoroughly replace, 
the traditional communication infrastructure, for the benefit of the affected
population as well as of first responders \cite{al2016stochastic}.

The rapid deployment of a communication network for data exchange between a
Rescue Command \& Control Center and ground units, or casualties, spread over the
territory, can be difficult \cite{manoj2007communication}. This is especially true in presence of natural obstacles
(mountains or valleys) that can limit the functionality of terrestrial
communication devices such as VHF radio equipment or line-of-sight (LOS)
data links. In these cases, UAVs (Unmanned Aerial Vehicles) represent a more
than valid solution for the rapid deployment of a secondary communication infrastructure \cite{erdelj2017help}.
Indeed, the UAV Data Link system can be equipped
with an additional broadcasting function, in order to directly
disseminate sensors and tactical data over a wide area, or connect
locally a small number of Ground Units. This can represent a valid
alternative to satellite communication equipment, which often suffers from the
lack of available channels and usually offers less channel bandwidth \cite{ibnkahla2004high}.
Additionally, satellite links are generally characterized by lower
resistance to attacks such as  jamming,
higher interceptability, and higher latency with respect to
wide band Data Links for UAVs.

In this paper, we address the provision of an alternative wireless coverage 
using UAVs, considering the specific use case of 
a population fleeing a disaster area riding their own vehicles. We  first 
review some literature work in \Sec{related_work}, while highlighting the novelty of our contribution. Our system model 
is described in \Sec{model}. Performance evaluation for 
the scenario illustrated in \Sec{refscen}, is shown against 
a baseline solution, in \Sec{results}. Finally, conclusions are drawn in \Sec{concl}.

\section{Related Work}
\label{sec:related_work}

We classify the related work in the following categories: i) UAV placement to provide wireless coverage, 
ii) UAV trajectory optimization, iii) exploitation of UAVs to sense information during disasters, and
iv) UAVs for disaster management. In the following, we provide more details about each of these categories.

\subsection{UAV placement for wireless coverage}

Here we analyze the works focusing on the placement of UAVs over a territory, in order to provide wireless coverage to users.
In this context, Mozaffari \textit{et al.} \cite{mozaffari2015drone} focus on the design, deployment, and performance analysis of UAV-based small Base Stations (BSs). They first develop an analytical framework to compute the optimal height for a single UAV-based BS in order to minimize the transmitted power. In the second part of their work, they study the optimal height and distance of two UAV-based BSs in order to maximize the coverage performance over the given areas. Bor \textit{et al.}  \cite{bor2016efficient} formulate the UAV placement problem in the 3D space with the goal of maximizing the revenue. Mozaffari \textit{et al.} \cite{mozaffari2016efficient} study the efficient deployment of multiple UAV-based BSs to provide wireless coverage for a set of users placed at ground level. They first analyze the coverage probability in the downlink direction (i.e., from the UAV-based BS to the user) as a function of the altitude and the antenna gain. Then they compute the placement of the UAVs in the 3D space with the aim to maximize the total coverage.

 Reina \textit{et al.} \cite{reina2018multi} design a genetic-based algorithm to solve the coverage problem of UAV-based networks. In particular, they model the deployment of a set of UAVs  as a wireless network design problem, which  optimizes the UAV positions. A similar problem to \cite{reina2018multi} is tackled by Zhao \textit{et al.} \cite{zhao2018deployment}, which propose  a set of algorithms for on-demand coverage while maintaining at the same time the connectivity between the UAVs.  Trotta \textit{et al.} \cite{trotta2018joint} propose a set of UAV deployment strategies in order to maximize geographical coverage, while considering UAV energy recharging operations at ground sites. In particular, they address a 3D placement problem, which is coupled with the scheduling of the UAVs recharging actions.

Although  the studies in  \cite{mozaffari2015drone,bor2016efficient,mozaffari2016efficient,reina2018multi,zhao2018deployment,trotta2018joint} demonstrate that the UAVs placement is an interesting and challenging problem and present sound solutions to it, 
our work substantially differs from previous work in several aspects, namely: i) the  dynamic selection of the areas to be covered, which depends on the evolution of the disaster over time; ii) the scheduling of the UAV moving, covering, and recharging actions; iii) the trajectory of the UAVs as a combination of zones that are visited over time; iv) the consideration of a realistic scenario based on the predicted mobility patterns of users.

\subsection{UAV trajectory optimization}

A second taxonomy of works deals with the problem of trajectory optimization of UAV-based wireless networks. 
Zhang \textit{et al.} \cite{zeng2016wireless} propose a basic UAV-based architecture and discuss the key design aspects. Among them, one important issue is the  UAV trajectory planning. Finding an optimal (or near optimal) solution to such problem is, in fact, beneficial to the distance between the user and the UAV and, consequently, to the users performance.  However, the authors state that retrieving the optimal path is very challenging, due to the presence of time-varying constraints, in terms, e.g., of connectivity, energy limitation, collision aspects, and terrain obstacles. Moreover, they also suggest to approximate the problem by dividing the time horizon into a set of discrete time slots. As a result, the trajectory of each UAV is given as a set of states that are visited over time. In line with them, this work considers: i) a discretized UAV path planning problem and ii) the time-varying constraints of UAV energy, UAV actions, and UAV-to-users connectivity.

Wu \textit{et al.} \cite{wu2018joint} consider a scenario where multi UAV-based BSs cooperatively serve a group of users placed at ground level. The considered problem takes into account the association of the users to the UAVs, their scheduling over time, the UAVs trajectories, and the UAVs transmit power. Unlike \cite{wu2018joint}, our work faces a different problem, where the dynamism of users, based on their movements over the disaster area, is introduced. Moreover, while the framework of \cite{wu2018joint} is tailored to a limited number of users,  in our work we focus on a large number of users, modeled following real-world data traces and a realistic scenario, and we account for the need of UAV recharging over time.

\subsection{UAVs for disaster sensing}

A third taxonomy of works targets the exploitation of UAVs for sensing operations during disasters. More in detail, Changchun \textit{et al.} \cite{changchun2010research} advocate the need of adopting UAVs for remote sensing and applications. Luo \textit{et al.} \cite{luo2015uav} design an application framework and a prototype of an UAV cloud-based architecture for disaster sensing.  However, their work is tailored to a video capture service, and not a wireless network service like in our case. 

Recently, Yanmaz \textit{et al.} \cite{yanmaz2018drone} introduce an UAV-based architecture for sensing operations, where each UAV is equipped with on-board sensors, as well as processing, coordination, and networking capabilities. The proposed architecture is implemented in a real testbed, which demonstrates its potentiality in terms of disaster assistance and aerial monitoring. With respect  to them, our work has a different objective, namely, the provisioning of the wireless service. Furthermore, we address some of the open issues left by \cite{yanmaz2018drone}, i.e., i) the coordinated path planning for a set of UAVs subject to dynamic goals (i.e., the data traffic coming from the users which are fleeing from the disaster area), and ii) the integration of the UAVs with a wireless service.

\subsection{UAVs for disaster management}

At last we consider the works analyzing the impact of UAV-based BSs that are deployed for  disaster management. In this context, Erdelj \textit{et al.} in their survey \cite{erdelj2016uav,erdelj2017help} analyze the studies dealing with this specific aspect. According to their taxonomy, our work falls inside the standalone communication systems category, where the UAVs are used to re-establish the communication infrastructure.
More in depth, Table~1 of \cite{erdelj2017help} reports the pros and cons of the fixed wing, rotary-wing (helicopter), and rotary-wing (quadcopter) UAV-based solutions. While fixed-wing UAVs are very effective in covering vast portions of territory, their total weight is limited. On the other hand, helicopter-based UAVs are able to carry even heavy loads. Finally, quadcopter UAVs are able to carry lower payload compared to the previous two solutions. However, the price of both fixed-wing and helicopters is much higher compared to the one of quadcopter UAVs. In line with them, in this work we consider UAV solutions based on quadcopters. Clearly, this choice has an impact on our model. For example, the authors of  \cite{erdelj2017help} point out the importance of automatically scheduling the UAV recharging actions. In our case, this aspect is carefully introduced in the formulation, as the flight time of quadcopter UAVs is pretty limited. Finally, the need of detailed problem formulations, able to provide an adequate service-level to users, is advocated by \cite{erdelj2017help}.

Merwaday \textit{et al.} \cite{merwaday2015uav} explore the adoption of UAV-based BSs for management of public safety communications during disasters. In particular, the gain in terms of throughput when placing the UAVs over the disaster area is analyzed. However, the movement of the UAVs over the disaster area is not considered. 

Bupe \textit{et al.} \cite{bupe2015relief} target the deployment of a set of UAVs in order to guarantee a first level of communication in a disaster area. The UAVs deployment and positioning is governed by a custom-based algorithm, which involves the creation of super-nodes that are in charge of running the heuristic.
Guevara \textit{et al.} \cite{guevara2015uav} assume that the UAVs are used to carry GSM Base Stations, and propose a solution based on the OpenBTS software. Although both the above studies \cite{bupe2015relief,guevara2015uav} are interesting steps towards a practical implementation of an UAV-based network, no results are shown.

Erdelj \textit{et al.} \cite{erdelj2017wireless} explore the use of Wireless Sensor Networks (WSNs) and UAVs for the management of disasters, by surveying the main works in the field. In particular, the main applications requiring WSNs and/or UAVs are classified according to the phase of the disaster. Moreover, Figure~5 in \cite{erdelj2017wireless} provides an example of a fixed UAV station. In line with them, in this work we assume that the UAVs are initially placed at fixed stations. Then, as soon as the disaster is detected, the UAVs are freely moved across the zones over time.

Finally, Zhang \textit{et al.} \cite{zhang2018analysis} explore the use of two UAVs providing coverage capabilities to a set of rescue vehicles. In particular, the coverage probability and the achievable data rate are analyzed. Differently from \cite{zhang2018analysis}, we consider a larger number of UAVs (i.e., more than two). Additionally, while in \cite{zhang2018analysis}  the communication from the UAV and the rest of the network is provided through satellite links, in our work we assume that this functionality is provided by a backhaul radio link between the UAV-based BS and the ground site.

\subsection{Novelty}

Summarizing, our novel contributions are as follows:
\begin{itemize}

\item we target user throughput in the set of areas covered by UAVs as the main performance metric and aim at ensuring a fair service to all users;

\item we pursue the aforementioned goal while scheduling the UAV actions such as  moving, recharging, and area covering, which leads an innovative optimization problem;

\item we tackle the complexity of a realistic scenario affected by a disaster, which involves more than 100,000 users, 500 areas, 20 UAVs and \textbf{150} time steps (each of them being 10 minute-long); 

\item we derive a methodology to efficiently deal with such a scenario and investigate the system performance.
\end{itemize}
To the best of our knowledge, none of the previous works has conducted a similar analysis.

\section{System model}
\label{sec:model}

This section describes our system model. We first deal with the mobility of UAVs  (\Sec{model-drones}) and then with its effect on the service quality (\Sec{model-quality}).

As a convention, in the following
\begin{itemize}
    \item calligraphic capital letters indicate sets, e.g.,~$\Ac$;
    \item Greek letters indicate variables, e.g.,~$\gamma(d,k,z)$;
    \item upper-case Latin letters indicate parameters, e.g.,~$B(d)$;
    \item lower-case Latin letters are used for indices.
\end{itemize}

\begin{table*}
\caption{Notation
\label{tab:notation}
\vspace{2mm}} 
\begin{tabularx}{1\textwidth}{|c|c|X|}
\hline
Symbol & Type & Meaning\\
\hline
\hline
$\Ac$ & Set & Area where vehicles can be\\
\hline
$\Dc$ & Set & UAVs\\
\hline
$\Kc$ & Set & Time steps\\
\hline
$\Lc\subseteq\Zc^2$ & Set & Pairs of nearby zones\\
\hline
$\Rc\subseteq\Zc$ & Set & Zones hosting recharge sites\\
\hline
\hline
$B(d)$ & Parameter & Battery capacity of UAV $d$ \\
\hline
$H$ & Parameter & Time horizon to consider when computing average throughput values\\
\hline
$N(a,k)$ & Parameter & Number of vehicles in area $a$ at step $k$\\
\hline
$T(a,z)$ & Parameter & Best-case throughput available to users in area $a$ from a UAV covering zone $z$\\
\hline
\hline
$\gamma(d,k,z)$ & Binary decision variable & Whether UAV $d$ covers zone $z$ at step $k$ or not\\
\hline
$\tau(d,k,z_1,z_2)$ & Binary decision variable & Whether UAV $d$ travels from zone $z_1$ to zone $z_2$ at step $k$ or not\\
\hline
$\rho(d,k,z)$ & Binary decision variable & Whether UAV $d$ recharges in zone $z$ at step $k$ or not \\
\hline
$\phi(a,d,k,z)$ & Real decision variable & Fraction of available spectrum resources assigned by UAV $d$ to area $a$ while covering zone $z$ at step $k$ \\
\hline
$\mu(a,k)$ & Real auxiliary variable & Per-vehicle, instantaneous throughput experienced by vehicles in area $a$ at step $k$\\
\hline
$\bar{\mu}(a,k)$ & Real auxiliary variable & Per-vehicle, average throughput experienced by vehicles in area $a$ between the steps $H-k$ and $k$\\
\hline

\end{tabularx}
\end{table*}

\subsection{UAVs mobility}
\label{sec:model-drones}

Throughout our paper, we refer to {\em zones} as the locations (in air) where drones can be, while {\em areas} are the locations (on the ground) they can serve. Specifically,
topology is described as a set~$\Ac$ of {\em areas} where vehicles in need of coverage can be located. We further identify a set~$\Zc$ of {\em zones} where a UAV can be stationed; a UAV located in one zone can cover vehicles in one or more areas, with a service quality that depends upon the distance between the UAV itself and the vehicles.

Pairs of zones that are close enough so that a UAV can travel from one to another in a time step are collected in set~$\Lc\subseteq\Zc^2$. Furthermore, some zones~$\Rc\subseteq\Zc$ host {\em recharge sites} where UAVs can
swap or, if the battery technology allows it, fast-charge, their batteries.
Time is divided into {\em steps}~$k\in\Kc$.

At every time step~$k$, each UAV~$d\in\Dc$ can perform one of the following three actions:
\begin{itemize}
    \item {\em cover} a zone~$z\in\Zc$, which is denoted through the binary variable~$\gamma(d,k,z)\in\{0,1\}$;
    \item {\em travel} from zone~$z_1$ to zone~$z_2$, such that~$(z_1,z_2)\in\Lc$, corresponding to the binary variable~$\tau(d,k,z_1,z_2)\in\{0,1\}$;
    \item {\em recharge} at zone~$z\in\Rc$, corresponding to the binary variable~$\rho(d,k,z)\in\{0,1\}$.
\end{itemize}
Setting the~$\gamma$, $\tau$ and $\rho$~variables is how we describe the movement of a UAV.

A first constraint we need to impose is that UAVs perform exactly one of the actions above in any given time step, i.e.,
\begin{multline}
\label{eq:oneaction}
\sum_{z\in\Zc}\gamma(d,k,z)+\sum_{z_1,z_2\in\Lc}\tau(d,k,z_1,z_2)+\\
+\sum_{z\in\Rc}\rho(d,k,z)=1, \quad\forall d\in\Dc,k\in\Kc.
\end{multline}
Also notice that \Eq{oneaction}  implies that UAVs that are recharging ($\rho$-variable set to one) do not cover any area ($\gamma$-variable set to zero), regardless of the location of recharge site.

UAVs move at most over one link in one time frame. It follows that UAVs cannot cover a zone~$z$ at a step~$k$ unless (i) they were covering the same zone at the previous time frame, or (ii) they were recharging there, or (iii) they just traveled there from another nearby zone~$z^\prime$: 
\begin{multline}
\label{eq:constr-cover}
\gamma(d,k,z)\leq \gamma(d,k-1,z)+\rho(d,k-1,z)+\\
+\sum_{z^\prime\in\Zc\colon (z^\prime,z)\in\Lc}\tau(d,k-1,z^\prime,z),\\
\quad\forall d\in\Dc,k\in\Kc,z\in\Zc.
\end{multline}

The same conditions must be met for a UAV to recharge at zone~$z\in\Rc$ at step~$k$:
\begin{multline}
\label{eq:constr-charge}
\rho(d,k,z)\leq \gamma(d,k-1,z)+\rho(d,k-1,z)+\\
+\sum_{z^\prime\in\Zc\colon (z^\prime,z)\in\Lc}\tau(d,k-1,z^\prime,z),\\
\quad\forall d\in\Dc,k\in\Kc,z\in\Rc.
\end{multline}

Furthermore, UAVs can travel over a link~$(z_1,z_2)\in\Lc$ at step~$k$ only if, at step~$k_1$, they were (i) covering or charging in~$z_1$ or (ii) traveling there from another zone~$z_3$:
\begin{multline}
\label{eq:constr-travel}
\tau(d,k,z_1,z_2)\leq \gamma(d,k-1,z_1)+\rho(d,k-1,z_1)+\\
+\sum_{z_3\in\Zc\colon (z_3,z_1)\in\Lc}\tau(d,k-1,z_3,z_1),\\
\quad\forall d\in\Dc,k\in\Kc,z_1,z_2\in\Lc.
\end{multline}

Last, we need to account for battery duration. Assuming that the same amount of battery is needed when traveling or covering, the battery capacity of UAV~$d$ can be expressed through a parameter~$B(d)$ corresponding to the maximum number of time steps between recharges. Ensuring that UAVs regularly charge is equivalent to saying that each UAV~$d$ must, in the last~$B(d)$ time steps, have recharged at least once:
\begin{equation}
\label{eq:charge}
\sum_{h=k-B(d)}^k \sum_{z\in\Rc}\rho(d,h,z)\geq 1,\quad
\forall d\in\Dc,k\in\Kc.
\end{equation}
\Eq{charge} models the fact that batteries can be recharged completely in one time step. As mentioned earlier, this can be achieved by swapping them with pre-charged ones.

\subsection{Coverage and service quality}
\label{sec:model-quality}

We are given the {\em best-case throughput} $T(a,z)$ with which a UAV staying in zone~$z\in\Zc$ can cover the vehicles in area~$a\in\Ac$. In order to account for the fact that a UAV in a certain zone can cover multiple areas, we introduce a real decision variable, $\phi(a,d,k,z)\in[0,1]$, expressing the fraction of spectrum resources that a UAV~$d$ covering zone~$z$ during step~$k$ devotes to covering area~$a$.
Without loss of generality, and motivated by the preponderance of downloads over uploads in real-world traffic~\cite{noi-tmc}, we only focus on user downstream traffic.
We then impose:
\begin{equation}
\label{eq:phi-cover}
\sum_{a\in\Ac} \phi(a,d,k,z)\leq \gamma(d,k,z),\quad\forall d\in\Dc,k\in\Kc,z\in\Zc.
\end{equation}
Eq.\,\Eq{phi-cover} ensures, at the same time, that (i) coverage of the vehicles staying in an area~$a$ can only be offered by UAVs located in nearby zones~$z$, i.e., such that~$T(a,z)>0$, and (ii) UAVs do not provide more capacity than they can. It is also interesting to notice how the capacity assignment across different areas can change over time.
As for interference between drones, we mandate that drones covering the same area use different radio resources::
\begin{equation}
\label{eq:phi-schedule}
\sum_{a\in\Ac}\sum_{d\in\Dc}\phi(a,d,k,z)\leq 1,\quad\forall k\in\Kc,z\in\Zc.
\end{equation}

If we know the number~$N(a,k)$ of vehicles staying in area~$a\in\Ac$ at step~$k\in\Kc$, then we can compute the average {\em per-user} throughput~$\mu(a,k)$, defined as:
\begin{multline}
\label{eq:mu}
\mu(a,k)=\frac{1}{N(a,k)}\sum_{z\in\Zc}T(a,z)\sum_{d\in\Dc}\phi(a,d,k,z),\\
\quad\forall a\in\Ac,k\in\Kc.
\end{multline}
It is worth pointing out how \Eq{mu} also accounts for the fact that multiple UAVs, possibly covering  different zones, can provide connectivity to the same area.

Finally, we  define our optimization objective. We need to (i) ensure fairness across different parts of the topology, i.e., different areas, but also (ii) exploit the fact that UAVs can move, hence the throughput can change over time. To this end, we define an auxiliary variable, $\bar{\mu}(a,k)$, expressing the {\em average} throughput experienced by area~$a$ in the last $H\geq 1$ frames, i.e.,
\begin{equation}
\label{eq:mubar}
\bar{\mu}(a,k)=\frac{1}{H}\sum_{h=k-H}^k \mu(a,h),\quad\forall a\in\Ac,k\in\Kc.
\end{equation}

Using the $\bar{\mu}$ variables, we define our objective as:
\begin{equation}
\label{eq:obj}
\max_{\gamma,\rho,\tau,\phi}\min_{a\in\Ac,k\in\Kc}\bar{\mu}(a,k).
\end{equation}
Importantly, \Eq{obj} combines the fairness coming from a max-min objective with the need to account for a longer time horizon, during which UAVs can roam across the topology and serve multiple areas.

\section{Reference scenario}
\label{sec:refscen}

We simulate a flooding in San Francisco and the traffic resulting from escaping cars using a combination of open data and state-of-the-art tools, as set forth next.

\subsection{Disaster}

In order to estimate which areas would be affected by a large-scale flooding, we use a software called Hazus~\cite{hazus}, developed by the American Federal Emergency Management Agency (FEMA) and able to simulate several types of disasters, including fires, earthquakes, and floodings. Given the scale of the disaster, e.g., the magnitude of an earthquake, Hazus is able to estimate to which extent each zone of the topology will be affected by the disaster itself, e.g., whether buildings therein will be partially or totally destroyed.

\begin{figure*}
\centering
\subfigure[\label{fig:fit}]{
\includegraphics[width=1\columnwidth]{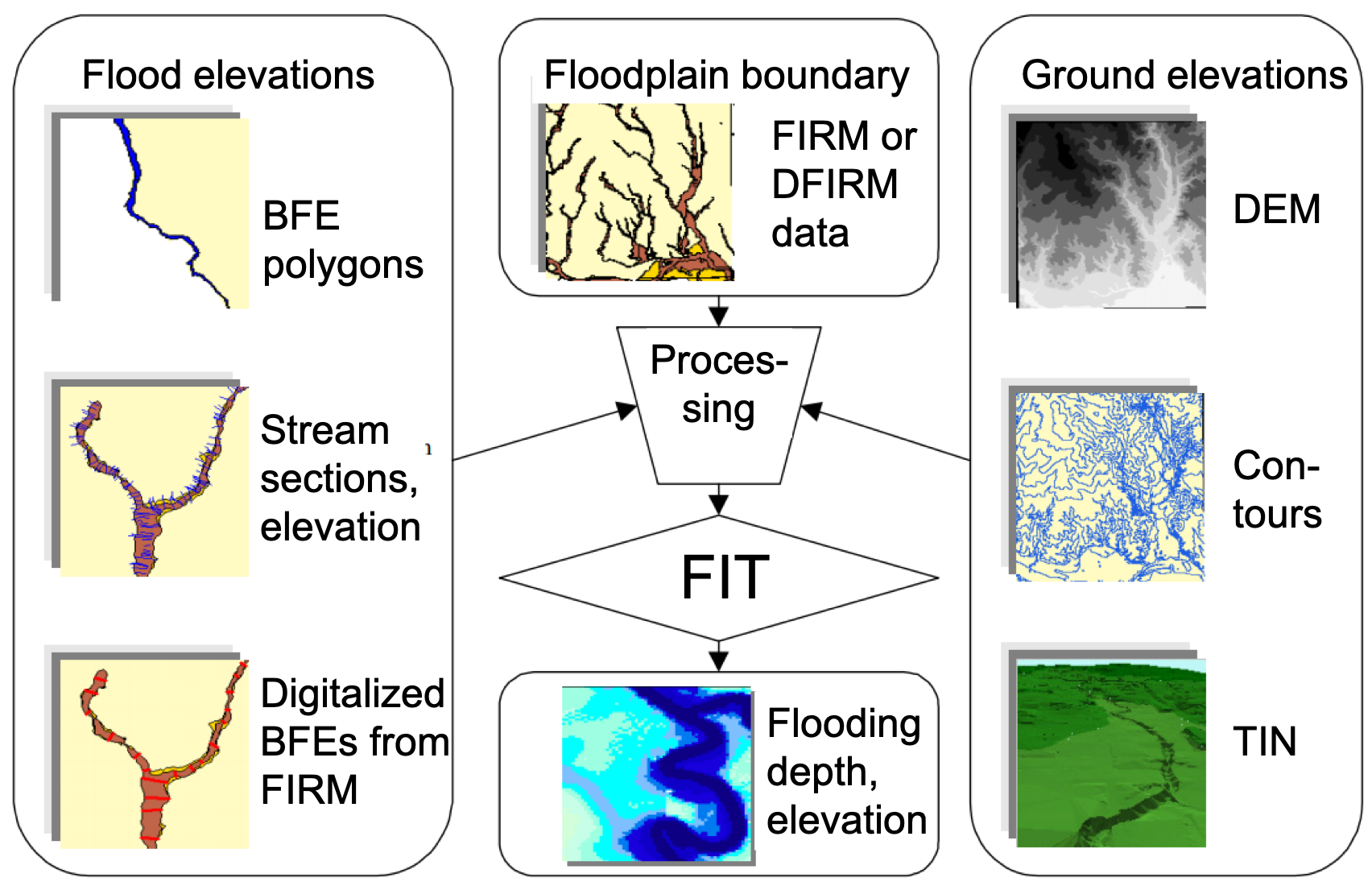}
} 
\centering
\subfigure[\label{fig:fit-levels}]{
\includegraphics[width=1\columnwidth]{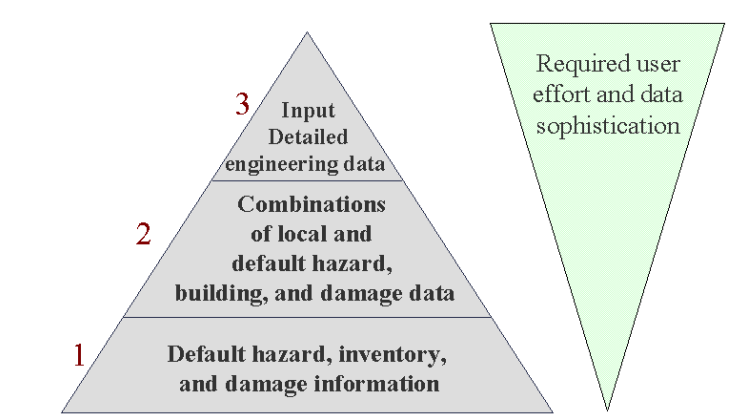}
} 
\caption{Basic architecture (a) and supported analysis levels (b) of the Flood Information Tool (FIT) used by Hazus. Source:~\cite{fit}.
} 
\end{figure*}

For floodings, Hazus leverages an ArcGIS extension called Flood Information Tool~\cite{fit}, whose basic architecture is summarized in \Fig{fit}. The FIT tool combines information about:
\begin{itemize}
    \item the expected flood elevation, in one of several supported formats (left-hand side of \Fig{fit});
    \item the extent of the floodplain in the area (top-center of \Fig{fit});
    \item the ground elevation of the areas that may potentially be affected by the disaster, in one of several supported formats (right-hand side of \Fig{fit}).
\end{itemize}

Using the above information, the FIT tool can perform one of three levels of analysis, summarized in \Fig{fit-levels}. If only basic information is supplied, it will return the extent of the areas affected by the disaster. If, on the other hand, the user provides more information about the type of area (urban, rural...), or the infrastructures therein, then the FIT tool will provide a detailed assessment of the damage caused by the flood and even its economic impact.

For our purposes, i.e., assessing the number of users that will be escaping from the disaster and the resulting vehicular traffic, a basic analysis (level~1 in \Fig{fit-levels}) is sufficient. The result is an ESRI shapefile representing the area affected by the disaster, as exemplified in \Fig{disaster}. We couple such information with population information from the US Census, as detailed next, to estimate the mobility of escaping users.

\subsection{Vehicular mobility}

Given the area hit by the disaster, we have to estimate (i) how many people will escape, (ii) when, and (iii) through which routes. We obtain the first information from official US census information~\cite{tiger}, which indicates how many people live in each census tract. As for the second, we estimate that all such people start evacuating as soon as the alarm is raised, i.e., they will all hit the road at the same time.

To simulate the trips of the escaping cars, we use an open-source traffic simulator called MATSim~\cite{matsim}. Started in 2006, the MATSim project has the goal of studying the traffic and congestion patterns resulting from the individual behaviors of vehicular users. It has been since extended to include such advanced features as schedule-based public transit, electric and autonomous cars, paratransit (special transportation for people with disabilities), and route re-planning as a consequence of current traffic conditions.

Since the inception of the project, efficiency has been a paramount goal of MATSim. To achieve this goal, its developers have used concepts and methodologies from such  domains as network science and particle physics: the models describing the behavior of individuals and vehicles are simplified to the essential, dropping computationally-expensive features that often have a minor impact on the final result.

\begin{figure}
\centering
\includegraphics[width=.9\columnwidth]{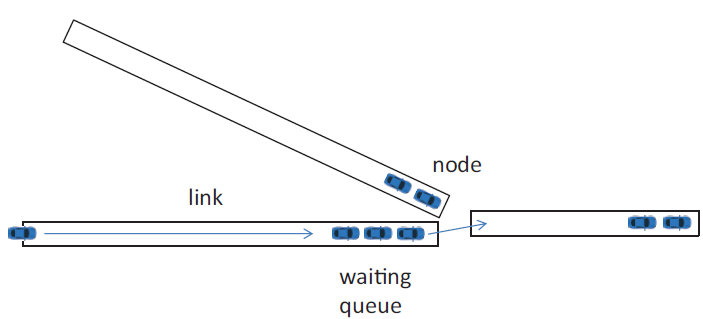}
\caption{The queue-based mobility model used in MATSim. Source:~\cite{matsim}.
\label{fig:matsim-mobility}
} 
\end{figure}

Following this philosophy, MATSim represents vehicular mobility through a queue-based model, exemplified in \Fig{matsim-mobility}. Each road segment corresponds to a queue; incoming vehicles are added to the end of the queue and stay in the segment until:
\begin{itemize}
    \item the free-flow travel time (i.e., the time it would take to travel that link at the maximum allowed speed) is elapsed, and
    \item there is no other vehicle in queue, and
    \item the next road segment is not congested, i.e., there is an available slot in the queue representing it.
\end{itemize}
Through this model, MATSim is able to account for road congestion and its effect on travel times; however, car-following dynamics and the position of individual vehicles {\em within} road segments are not captured,
In this respect, MATSim differs from more popular, microscopic mobility simulators like SUMO~\cite{sumo} that model the decisions of individual drivers, e.g., the distance they keep from the preceding car.

Such a {\em mesoscopic} approach yields much faster running times than SUMO
and thus the ability to cope with larger-scale scenarios. Furthermore, it perfectly fits our needs, as road segments correspond to areas in set~$\Ac$, and the number of vehicles therein correspond to the $N(a,k)$~parameters.

\begin{figure}
\centering
\includegraphics[width=1.0\columnwidth]{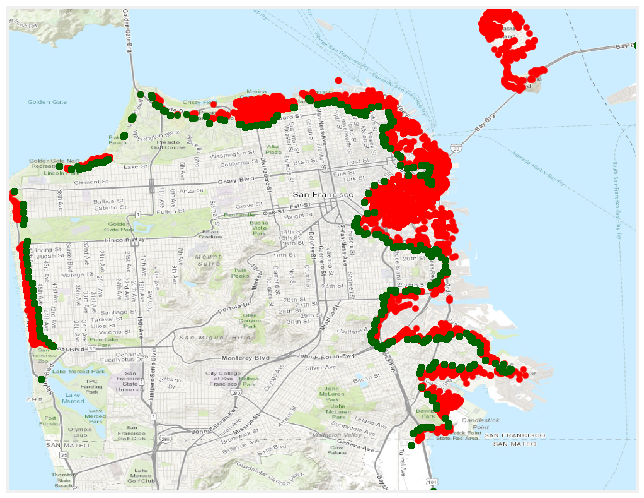}
\caption{
Areas affected by the flooding from where vehicles escape (red dots), and safe areas vehicles flee to (green dots).
    \label{fig:disaster}
} 
\end{figure}

\begin{figure*}
\centering
\subfigure[\label{fig:areas}]{
\includegraphics[width=1.2\columnwidth]{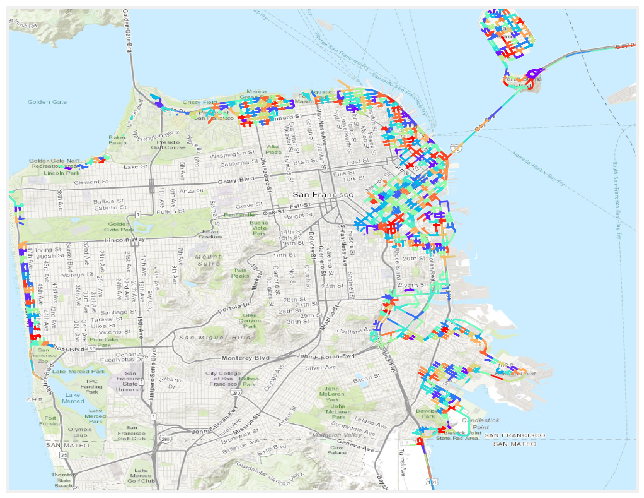}
} 
\hspace{1cm}
\subfigure[\label{fig:zones}]{
\includegraphics[width=1.2\columnwidth]{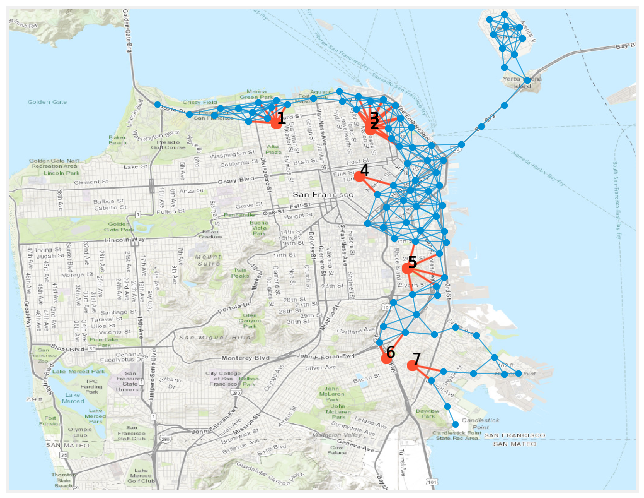}
} 
\caption{
Areas (a) and zones (b) in the reference scenario.
In (a), different colors correspond to different areas. In (b), blue dots represent zones and blue segments connect pairs of zones between which drones can travel in one time step, while red dots correspond to recharge sites and red segments link them with the zones reachable from them in one time step.
\label{fig:refscen}
} 
\end{figure*}

\begin{figure*}
\centering
\subfigure[\label{fig:spent-time}]{
\includegraphics[width=.9\columnwidth]{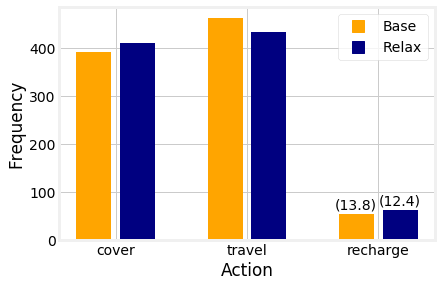}
} 
\hspace{1cm}
\subfigure[\label{fig:mission-duration}]{
\includegraphics[width=.9\columnwidth]{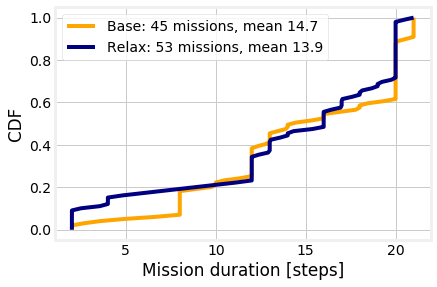}
} 
\caption{
Actions performed by UAVs under the baseline (``Base'') and relaxation-based (``Relax'') strategies (a); duration of missions (b). In (a), the numbers over the bars for the ``recharge'' action indicate the average number of battery units replenished by UAVs at each charge.
} 
\end{figure*}

\subsection{Network capacity and topology}

Lastly, we have to estimate the throughput between vehicles in certain areas and UAVs covering them from a certain zone, i.e., the $T(a,z)$~parameters. To this end, we perform the following three steps~\cite[Sec.~6]{noi-tmc}.
First, we compute the attenuation between any zone/area pair, using the ITU-recommended model~\cite{itu} for micro-cells in line-of-sight (LOS) conditions:
\begin{equation}
\label{eq:att}
PL_{dB}{=}40\log d{+}7.8{-}18\log h_{BS}{-}18\log h_{UE}{+}2\log f
\end{equation}
In \Eq{att}, $PL$ is the path loss (in dB), $d$ is the distance between zone and area, $h_{BS}$ is the height of the drone, which we assume to be 50~m, and~$h_{UE}$ is the height of the users, which we assume to be 1.5~m.  Furthermore, $f$ is the frequency, which we assume to be 1.8~GHz, i.e., the same frequency used by LTE micro-cells. This reflects the popular~\cite{r1} notion that drones will act as mobile micro-cells when providing cellular coverage.

As a second step, we compute the power received from any source/destination (i.e., zone/area) pair. For drones, we assume a 30~dBm transmission power, again matching that of LTE micro-cells. Then, considering a thermal noise of -121.45~dBm, we compute the signal-to-noise ratio (SNR) experienced by each zone/area pair.

Finally, using the experimental data in~\cite{sinr2rb}, we map SNR values into the amount of data each resource block can transport. In LTE, resource blocks are the atomic units of radio resources -- consisting of a combination of time, frequency, and code -- that a transmission-receiver pair can be assigned. Experimental studies such as~\cite{sinr2rb} establish how much data can be transferred in a single resource block, given the SNR; this, along with the number of resource block assigned to each source-destination pair, allows us to compute the network throughput.
Notice that we compute the SNR instead of the signal-to-noise-and-interference ratio (SINR), which corresponds to neglecting interference from the regular cellular networks -- a fair assumption since the very purpose of sending UAVs to an area is to make up for a disabled or damaged cellular network. 
Hence, no interference is to be expected from regular base stations; as for interference from other drones, it is prevented by constraint \Eq{phi-schedule}.

We derive our numerical results with reference to the topology in \Fig{disaster}. Using k-means clustering, we divide the topology into 500 areas, depicted in \Fig{areas}, which compose the set~$\Ac$
and whose average size is equal to~500~m$^2$.
 We then identify 100 zones, 7 of which also host recharge sites, corresponding to sets~$\Zc$ and~$\Rc$ in the system model and represented by blue and red dots, respectively, in \Fig{zones}. Furthermore, \Fig{zones} reports links in~$\Lc$, drawn between any two zones closer than 1~km. Time steps in~$\Kc$ correspond to 10-minute time intervals.
 
\subsection{UAVs strategies}

Solving the problem presented in \Sec{model} to optimality would yield the best possible actions that every UAV should perform at any given time. However, directly solving such a problem is impractical in medium- to large-scale scenarios, owing to its complexity. We therefore solve a {\em relaxed} version of the problem, where the binary variables~$\gamma,\tau,\rho\in\{0,1\}$ are replaced by their continuous counterparts~$\tilde{\gamma},\tilde{\tau},\tilde{\rho}\in[0,1]$.

The values of the $\tilde{\gamma},\tilde{\tau},\tilde{\rho}$~variables cannot be directly used to steer a UAV; rather, for every time step, we perform the action associated with the highest relaxed variable. If, as an example, at time step~$k$ the variables associated with UAV~$d$ are~$\tilde{\gamma}(d,k,z_1)=0.3$ and~$\tilde{\tau}(d,k,z_1,z_2)=0.7$, UAV~$d$ will travel to~$z_2$, as if it were~$\tau(d,k,z_1,z_2)=1$.

We compare against a baseline solution whereby UAVs choose as their waypoint the least-recently visited zone of the topology and travel there through the shortest path, covering each zone they traverse for one time step before proceeding. When the waypoint is reached, a new one is selected. If a UAV's course would bring it so far from the closest charging site that its battery level would be insufficient to reach it, the UAV changes its course and heads to the charging site.

\section{Numerical results}
\label{sec:results}

We carry out our numerical evaluation taking as a reference the flooding depicted in \Fig{disaster}, affecting a total of 172,000~people. We assume that the cellular network is utterly disabled in affected areas, seek to provide coverage to the users therein through a total of 20~drones, whose battery lasts for 20~time steps. The time horizon used when computing objective metrics is~$H=30~\text{steps}$. When drones visit a recharge site, their batteries are swapped with fully-charged ones, hence, drones departing from sites always have a battery level equal to~$B(d)$.

In this section, we characterize the mobility of UAVs under the baseline and relaxation-based strategy 
(\Sec{sub-mobility}), as well as the coverage they are able to provide to the evacuating vehicles (\Sec{sub-coverage}).

\subsection{UAV mobility}
\label{sec:sub-mobility}

A first aspect we are interested in is how UAVs spend their time, i.e., how often they cover one zone (``cover''), travel between zones (``travel''), or recharge at a site (``recharge''). As summarized in \Fig{spent-time}, the ``cover'' and ``travel'' actions account for the majority of UAV time. It is also interesting to notice that those two actions have roughly equal frequency, i.e., UAVs tend to cover a zone for one time step before moving to a different one. The relaxation-based strategy is associated with more covering and less traveling, i.e., drones stay longer in the same zone if needed, e.g., if they can cover multiple areas from it. It is also interesting to notice the frequency of the ``recharge'' actions: under the relaxation-based strategy, drones recharge more frequently and for smaller amounts. Again, this is a consequence of the greater flexibility of the scheme, which allows UAVs passing near a recharge site to replenish their battery even if it is not (yet) completely drained.

\Fig{mission-duration} depicts the number and duration of the missions performed by UAVs under the two strategies. We can observe that, under the relaxation-based strategy, UAVs perform a higher number of missions, thus reaching more areas of the topology. Clearly, such missions are, on average, shorter than under the baseline strategy. However, looking at the CDF, it is interesting to see that the baseline strategy is associated with a fairly large number of very short missions: those missions are cut short because of the need to recharge -- a condition that, as discussed earlier, is avoided under the relaxation-based strategy.

\subsection{UAV coverage}
\label{sec:sub-coverage}

We are interested in two complementary aspects of the coverage provided by UAVs to evacuating vehicles: the overall throughput and the fairness across different areas.
Recall that all traffic and throughput values only concern downlink traffic.

\begin{figure}
\centering
\includegraphics[width=.9\columnwidth]{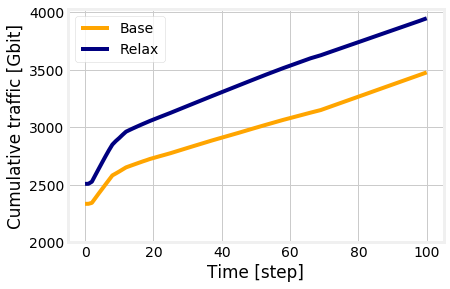}
\caption{
Total throughput for the baseline (``Base'') and relaxation-based (``Relax'') strategies.
    \label{fig:thp}
} 
\end{figure}

\begin{figure*}
\centering
\subfigure[\label{fig:shades-base}]{
\includegraphics[width=1.2\columnwidth]{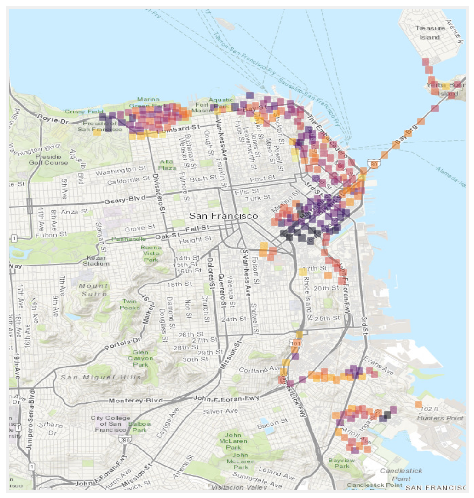}
} 
\hspace{1cm}
\subfigure[\label{fig:redgreen}]{
\includegraphics[width=1.2\columnwidth]{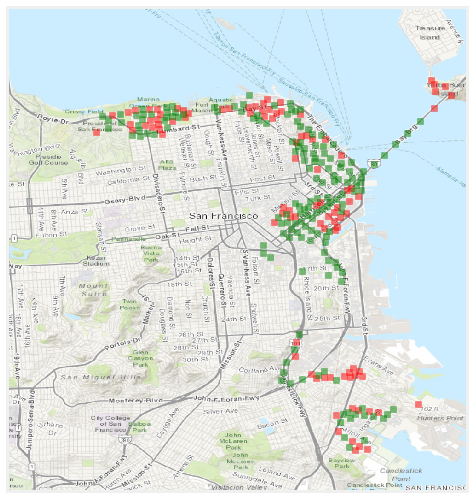}
} 
\caption{
Per-area throughput under the baseline strategy, with lighter colors corresponding to higher throughput (a); effect of moving to the relaxation-based strategy, with red and green shades corresponding to, respectively, worse and better throughput values.
} 
\end{figure*}

\Fig{thp} shows the overall throughput for the baseline and relaxation-based strategies, with the latter providing substantially and consistently better performance. This is due to the fact that the optimization problem accounts for additional information with respect to the baseline strategy, including the number of vehicles present in each area and the throughput with which they can be served.

In \Fig{shades-base}, we summarize the throughput experienced by vehicles in individual areas under the baseline situation; lighter colors correspond to better throughput. We can observe that some areas experience a better throughput than others, e.g., the Marina district in the North. Moving to \Fig{redgreen}, depicting the effects of switching to the relaxation-based strategy, we can observe that (i) most areas benefit from such a change and (ii) more importantly, the throughput of disadvantaged areas such as those along Market street is substantially increased.

In other words, thanks to the objective \Eq{obj}, the relaxation-based strategy results in a better fairness. This is confirmed by the values of the Jain's fairness index~\cite{jains}, computed considering the total throughput in different areas: 0.34~for the baseline strategy, and 0.40~for the relaxation-based one.

\section{Conclusions}

We have addressed the deployment of UAVs in disaster scenarios, with the aim to replace disrupted cellular infrastructure and bear the surge of user traffic typical of emergency situations. To this end, we defined an optimization problem in the context of UAV deployment in areas affected 
by natural disasters. Our goal was to determine the best UAV coverage that maximizes user throughput,
in a fair manner across different parts of the topology. Our numerical simulations on a real
map of the San Francisco area, considering realistic UAV mission parameters confirmed that
our solution provides a higher throughput and a better fairness when compared to a 
baseline solution whereby a UAV flies in a beeline to the least-recently visited zone of 
the topology, until a low battery forces it to return to base.

\label{sec:concl}

\section*{Acknowledgments}

This work  was supported by the European Commission through the I-REACT project  (grant agreement no.\,700256), and by the University of Rome Tor Vergata project BRIGHT (Call Mission Sustainability).

\section*{References}
\bibliographystyle{IEEEtran}
\bibliography{refs}

\end{document}